\documentclass[reprint,amsmath,amssymb, aps, prl]{revtex4-1}
\UseRawInputEncoding
\usepackage{graphicx}
\usepackage{siunitx}
\usepackage[colorlinks=true, citecolor=blue, urlcolor=blue]{hyperref}
\usepackage{orcidlink}
\usepackage{upgreek}
\usepackage{mathrsfs}

\newcommand{\dagg}{^\dagger}
\newcommand{\ket}[1]{|#1\rangle}
\newcommand{\bra}[1]{\langle#1|}
\renewcommand{\-}[1]{\mathrm{#1}}

\renewcommand{\:}[1]{\tilde{#1}}

\begin{document}
\preprint{APS/123-QED}
\title{Suppression of Local Decay in non-Markovian Waveguide QED}
\author{Yuan Liu$^1$\orcidlink{0009-0006-7349-0903}}
\author{Linhan Lin$^1$\orcidlink{0000-0003-0556-9143}}
\email{linlh2019@mail.tsinghua.edu.cn}
\author{Hong-Bo Sun$^{1, 2}$\orcidlink{0000-0003-2127-8610}}
\email{hbsun@tsinghua.edu.cn}
\affiliation{%
 $^1$State Key Laboratory of Precision Measurement Technology and Instruments, Department of Precision Instrument, Tsinghua University, Beijing 100084, China\\
 $^2$State Key Laboratory of Integrated Optoelectronics, College of Electronic Science and Engineering, Jilin University, Changchun 130012, China
}
\begin{abstract}
Atoms coupled to the same environment interfere with each other to yield super- or sub-radiance. Specifically, atoms in subradiant states are promising candidates for long-lifetime qubits and quantum memory because of the immunity to the common environment. However, subradiant states can still be influenced by local environments, which are incoherent for different atoms and cannot be canceled out through interference. Here we propose to break this limit by preparing a waveguide QED system in the non-Markovian regime, where the ultra-small decay rate arises because of the retarded interaction.  We further show that similar effect occurs spontaneously by self-interference and can be stressed by cooperative coupling. 
\end{abstract}
\maketitle

\textit{Introduction.--}Spontaneous emission roots in the quantum nature of the electromagnetic (EM) field \cite{Scully_Zubairy_1997}. For multiple atoms coupled to the common EM modes (e.g., EM modes of a waveguide or an optical cavity), the emission processes of different atoms are not independent, but interfere with each other to yield super- or sub-radiance when the interference is constructive or destructive, respectively \cite{Sheremet23RMP, Reitz22PRXQuantum, Evans18S, Arjan13S, Tiranov23S, Tudela15Oct, Norcia16SA}. Specifically, atoms in subradiant states are promising candidates for long-lifetime qubits or quantum memory \cite{Sangouard11RMP, Suter16RMP, Zanardi97Oct, Facchinetti16Dec, Garcia17PRX}, and can also be useful for quantum metrology \cite{Orioli22PRX, Zhang23Nov, DeMille24NPhys}.

In atom arrays with deep sub-wavelength inter-atom distances, the environment of neighbour atoms is coherent and there exists subradiant states with spontaneous emission rate much lower than $\gamma_0$, the so called Einstein's $A$ coefficient \cite{Rui20N, Guimond19Mar, Shahmoon17Mar, Hebenstreit17Apr, Liu23Chip, Orioli19Nov, Zhang20Dec}. However, sub-wavelength inter-atom distances make the individual addressing of atoms difficult and more flexible platforms are desired \cite{Guerin16Feb, Ferioli21PRX, Takasu12Apr}. Recently, waveguide quantum electrodynamics (QED) has emerged as a research realm ideally suited for investigating the collective light-atom interactions because of the infinitely long atom-atom interaction length and much stronger atom-photon interaction strength than in free space \cite{Sheremet23RMP, Roy17RMP, Goban15Aug, Pennetta22Feb, Liedl23Apr, Lechner23Sep, Tabares23Aug, Liedl24PRX, Shi24arXiv, Solano17NC, Corzo19N, Sipahigil16S}. Though immune to the common environment (i.e., the waveguide EM modes), subradiant states in waveguide QED can still be influenced by local environments, which are incoherent for different atoms when inter-atom distances are larger than the transition wavelength and the induced decay cannot be cancelled out through interference \cite{DeVoe96Mar, Masson20Dec, Cipris21Mar, Holzinger22Dec, Salhov24May}, setting a lower bound of the total decay rate $\gamma\gtrsim\gamma_0$.

Here we propose to break this lower bound by preparing a waveguide QED system in the non-Markovian regime \cite{Dinc19PRRes, Ask22Feb, ZhangYX23Nov, Gaikwad24May, Cilluffo24arXiv}, where the total decay rate $\gamma<\gamma_0$ stems from the retarded interaction \cite{Sinha20Jan, Carmele20PRRes, Regidor21PRRes, Qiao22Aug}. A photon emitted by one atom propagates for a nonnegligible time $T$ before interacting with the other atom. Energy quanta are partly stored in the propagating photons, making the system partly immune from local atom emission into the free space. The interference with another atom protect the propagating photons from being coupled to the scattering modes of the waveguide. Thus, the total decay rate can be lower than $\gamma_0$. We further show that similar effect arises in a system with time-delayed feedback, where the atom is fed by its historical version and the ultra-small decay rate originates from self-interference. The atom can be replaced by a collection of atoms, where the cooperative emission improves the atom-waveguide coupling efficiency, thus magnifying the local-decay suppression effect.

\begin{figure}
    \centering
    \includegraphics[width=\columnwidth]{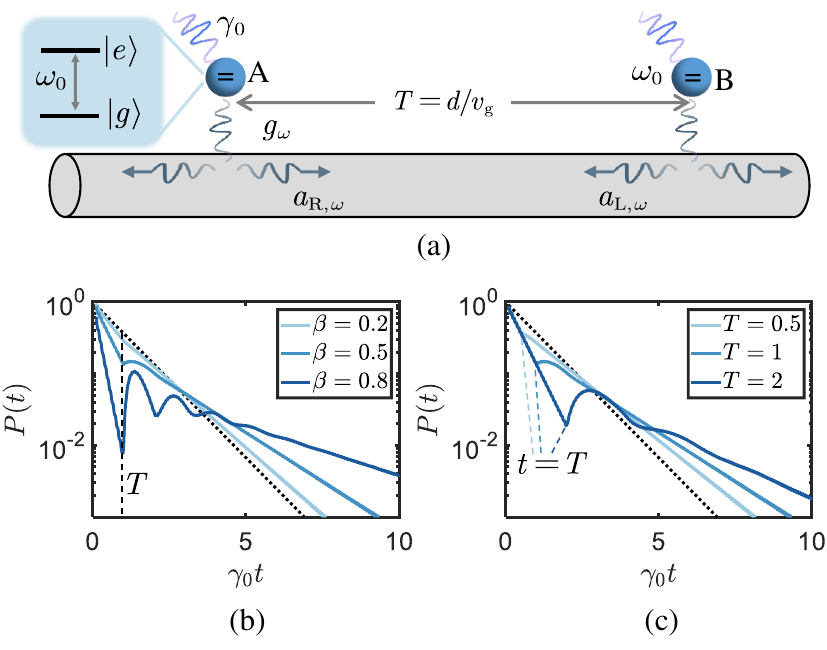}
    \caption{(a)~Setup and coupling scheme. Two identical two-level atoms A and B with distance $d$ are coupled to a common 1D waveguide. Atomic excitation probability $P(t)$ versus time (in units of $1/\gamma_0$) in semi-logarithmic coordinate (b)~for different coupling efficiencies $\beta$ and (c)~for different inter-atom temporal distances $T$. We set $T=1$ in (b) and $\beta=0.5$ in (c), respectively, and $\gamma_0=1$ is kept unchanged in all cases. The black dotted line for exponential decay $P(t)=\exp(-\gamma_0t)$ is plotted for comparison.}
    \label{Sub1}
\end{figure}

\textit{Model.--} As schematically illustrated in Fig.~\ref{Sub1}(a), two identical atoms A and B with a central transition frequency $\omega_0$ are coupled to a common one-dimensional (1D) waveguide with distance $d$. The atoms are described by the Pauli operators $\sigma_i=\ket{g}_{ii}\bra{e}$ and $\sigma_i^+=\ket{e}_{ii}\bra{g}$, where $\ket{g}_i$ and $\ket{e}_i$ are the ground and excited states of atom $i$ for $i\in\{\-{A,B}\}$. The spatial distance $d$ thus set a temporal distance $T=d/v_\-g$ for the atoms, i.e., the influence of one atom on the other is retarded by $T$. Here $v_\-g$ is the group velocity of a wave packet with central frequency $\omega_0$. The waveguide EM modes at frequency $\omega$ are described by the annihilation (ceation) operators $a_{\-R,\omega}$ $(a_{\-R,\omega}\dagg)$ and $a_{\-L,\omega}$ $(a_{\-L,\omega}\dagg)$ for the right- and left-propagating photons, respectively. The 1D waveguide is assumed to be bidirectional, i.e., each atom couples to the waveguide EM mode of frequency $\omega$ with strength $g_\omega$ regardless of the propagation direction. 

With the clear description of the atom-waveguide coupling, we obtain the full Hamiltonian $H=H_0+H_\-{int}$ governing the dynamics of the system, where ($\hbar=1$)
\begin{equation}
    H_0=\omega_0\sum_{i\in\{\-{A,B}\}}\sigma_i^+\sigma_i+\int_0^\infty\-d\omega\,\omega\left(a_{\-R,\omega}\dagg a_{\-R,\omega}+a_{\-L,\omega}\dagg a_{\-L,\omega}\right),
\end{equation}
\begin{equation}
    \begin{aligned}
        H_\-{int}={}&\sum_{i\in\{\-{A,B}\}}\int_0^\infty\-d\omega\left(g_\omega\sigma_i^+a_{\-R,\omega}\-e^{\-ik_\omega x_i}\right.\\
        &+\left.g_\omega\sigma_i^+a_{\-L,\omega}\-e^{-\-ik_\omega x_i}+\-{H.c.}\right).
    \end{aligned}
\end{equation}
At time $t=0$, the waveguide is assumed to be in the vacuum state such that $\langle a_{\-R,\omega}\dagg a_{\-R,\omega}\rangle=\langle a_{\-L,\omega}\dagg a_{\-L,\omega}\rangle=0$, and the atoms are assumed to be in the antisymmetric superposition state (the so called \emph{dark state} \cite{Evans18S, Holzinger22Dec}) $\ket{D}=(\ket{eg}-\ket{ge})/\sqrt{2}$. For $t>0$, the atoms are coupled to the waveguide and the time-evolving state can be written as 
\begin{align}
    \ket{\psi(t)}={}&\sum_{i\in\{\-{A,B}\}}c_i(t)\sigma_i^+\ket{G,\varnothing}\\
    &+\int_0^\infty\-d\omega\left[c_{\-R,\omega}(t)a_{\-R,\omega}\dagg+c_{\-L,\omega}(t)a_{\-L,\omega}\dagg\right]\ket{G,\varnothing},\nonumber
\end{align}
where $c_i(t)$ is the probability amplitude of atom $i$ in excited state and $c_{\-R/\-L,\omega}(t)$ the amplitude of the right-/left-propagating waveguide mode with frequency $\omega$. $\ket{G,\varnothing}$ denotes the state where all atoms are in the ground state and the EM field is in the vacuum state.
  
After tracing off the degrees of freedom of the waveguide, and considering the free space EM modes induced local decay rate $\gamma_0$, the equation of motion for atom $i\in\{\-{A,B}\}$ is derived to be \cite{SM}
  \begin{equation}\label{ODE_ci}
      \dot{c}_i(t)=-\frac{\gamma}{2}\left[c_i(t)+\beta c_{j\neq i}(t-T)\-e^{\-i\varphi}\Theta(t-T)\right].
  \end{equation}
Here $\gamma=\gamma_0+\gamma_{1\-D}$ is the total decay rate of an individual atom, with $\gamma_{1\-D}=2\int_0^\infty|g_\omega|^2\-d\omega$ the waveguide EM modes induced decay rate and $\beta=\gamma_{1\-D}/\gamma$ the coupling efficiency. $\varphi=\omega_0|x_\-A-x_\-B|/v_\-g$ is the photon propagation phase between sites A and B. In what follows we assume that the atoms are positioned such that $\-e^{\-i\varphi}=0$, which guaranties that the dark state $\ket{D}$ coincides with the subradiant state in the Markovian regime if the retardation $T$ is ignorable \cite{Tiranov23S, Arjan13S}. The Heaviside step function $\Theta(t-T)$ shows explicitly that the influence of one atom on the other is retarded by $T$, before which the atoms decay exponentially as if there were only one atom.

\textit{Results.--} The time evolution of the dark state $\ket{D}$ can be obtained by solving Eq.~\eqref{ODE_ci} analytically, as detailed in the Supplemental Material \cite{SM}. In Fig.~\ref{Sub1}(b)(c), the total excited state population $P(t)=\sum_{i\in{\-A,\-B}}|c_i(t)|^2$ is shown under different coupling efficiencies $\beta$ of the atoms to the waveguide and for different retardation $T$. To capture the main feature of the dark state evolution, we introduce the excited-state decay rate $\Gamma(t):=-(\-d/\-dt)\ln P(t)$. When $t<T$, the atoms decay exponentially with the same decay rate of an individual atom coupled to the waveguide, i.e., $\Gamma(t)=\gamma$, which can also be directly seen from Eq.~\eqref{ODE_ci}. At $t=T$, the photon emitted by one atom begins to arrive at the other atom, and the population $P(t)$ grows and oscillates because of photon reabsorption. When $t>T$, the population evolution approaches exponential decay, but astonishingly, with a decay rate even smaller than $\gamma_0$.  To sum up, we have $\Gamma(t)=\gamma$ for $t<T$ and $\Gamma(t)<\gamma_0$ for $t>T\wedge t\nsim T$.

As we fix the decay rate into the free space $\gamma_0$ unchanged, a higher coupling efficiency $\beta$ means a higher decay rate into the waveguide and hence a higher total decay rate $\gamma$, as evidenced in Fig.~\ref{Sub1}(b) when $t<T$. However, when $t>T$, the population $P(t)$ decays \emph{slower} with larger $\beta$. In Fig.~1(c), it is noticed that a larger retardation $T$ also leads to a smaller decay rate when $t>T$, even though any \emph{local} coupling parameter remains unchanged. Specifically, in the case that the retardation $T$ is small, we obtain \cite{SM}
\begin{equation}\label{DecayRate_D}
    \Gamma\simeq\frac{1}{1+\gamma_{1\-D}T/2}\gamma_0.
\end{equation}
Although this expression is mathematically solid only when $T$ is small, it captures the main features of $\Gamma$ when $t>T$ that $\Gamma$ decreases monotonically with both $\beta$ (and consequently $\gamma_{1\-D}$) and $T$. When $T\to0$, we have $\Gamma\to\gamma_0$, as is the result of the well-studied Markovian situation.

\begin{figure}
    \centering
    \includegraphics[width=\columnwidth]{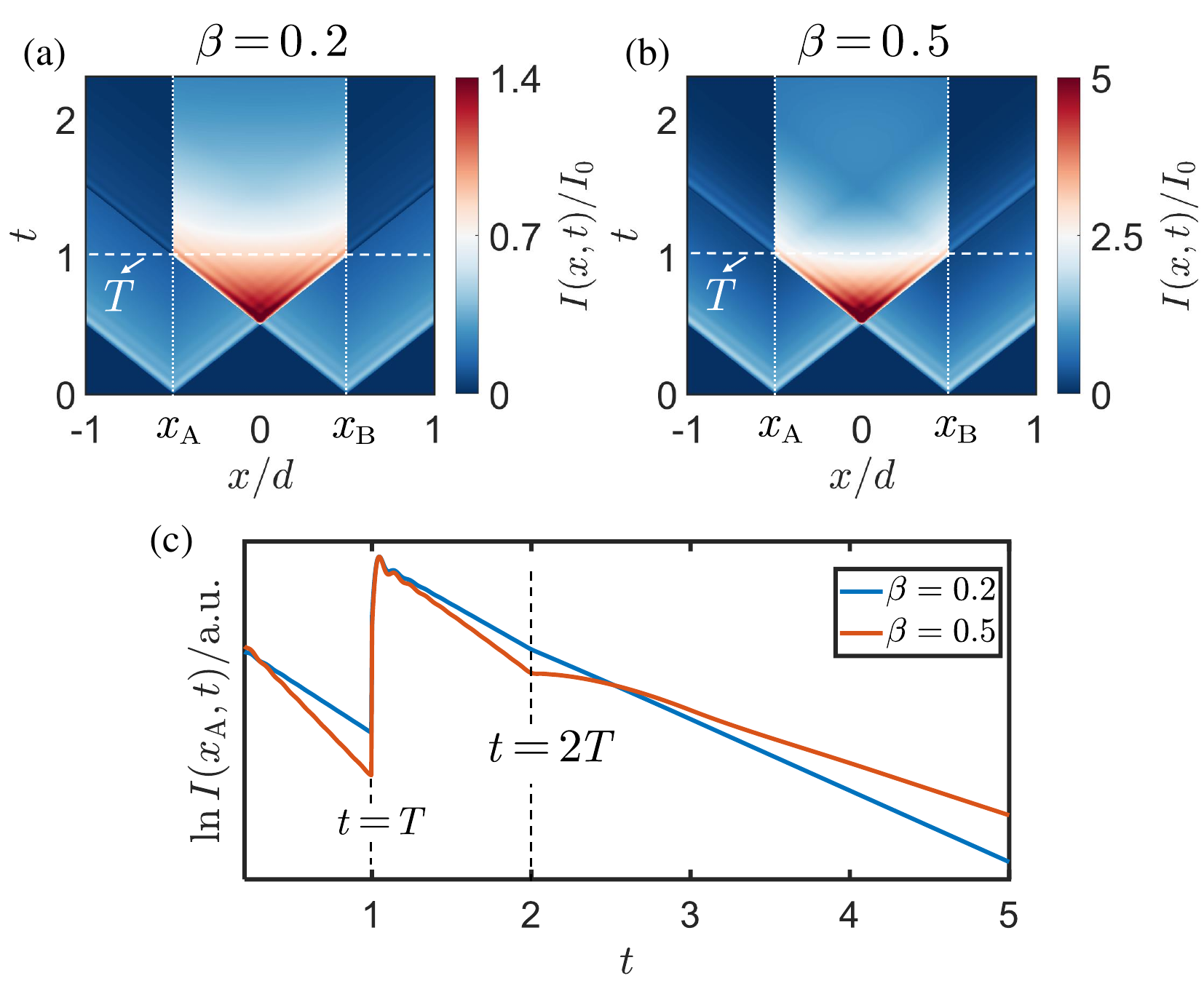}
    \caption{ EM field intensity as a function of position and time for (a) $\beta=0.2$ and (b) $\beta=0.5$, respectively. Atom A and atom B are positioned at $x_\mathrm{A}$ and $x_\mathrm{B}$, respectively, and are marked by the vertical dotted lines. Time $t=T$ is marked by horizontal dashed lines. We set $T = 1$ and $\gamma_0 = 1$ in both cases. Pay attention to the different colour bars. (c) Normalized EM field intensity at $x=x_\mathrm{A}+0^+$. Times $t=T$ and $t=2T$ are marked by vertical dashed lines.}
    \label{Sub2}
\end{figure}

To gain insight into this non-Markovian local-decay suppression effect, we calculated the EM field intensity $I(x,t)$ in \cite{SM}, and the results for different $\beta$ are presented in Fig.~\ref{Sub2}. Let us focus on Fig.~\ref{Sub2}(a)(b) first. When $t\in(0,T/2)$, the atoms at $x_\-A$ and $x_\-B$ emit independently, from where the EM fields propagate bidirectionally. At $t=T/2$, EM fields originating from $x_\-A$ and $x_\-B$ begin to interfere with each other. When $t=T$, the EM field emitted by one atom arrives at the other atom and photon reabsorption occurs, in consistence with the results in Fig.~\ref{Sub1}(b)(c). It is noted that the EM fields emitted when $t>T$ are mainly localized in $x\in(x_\-A,x_\-B)$, outside which the EM field vanishes because of the destructive interference between the EM fields emitted by different atoms. 

Actually, the steady state of our system corresponds to an atom-photon bound state (BS) if $\gamma_0=0$ \cite{Calajo19Feb, Sinha20Jan, Scigliuzzo22PRX}, mimicking the well-studied bound states in the continuum in optics \cite{Hsu16NatRevMater}. When $\gamma_0\neq0$, the steady state at $t>T$ is not perfectly bounded, and thus is a quasi-BS. The energy of this quasi-BS is partly confined in the EM field in $x\in(x_\-A,x_\-B)$, which is totally free from dissipation if the propagation loss is negligible. The energy stored in the excited atoms, on the other hand, suffer from local spontaneous decay with rate $\gamma_0$. However, the local dissipation of the atoms is continuously compensated by the EM field, thus leading to a ultra-small excited state decay rate $\Gamma<\gamma_0$. As is illustrated in Fig.~\ref{Sub2}(a)(b): when $t<T$, the larger the coupling efficiency $\beta$, the more energy emitted into the EM field in  $x\in(x_\-A,x_\-B)$; when $t>T$, the field intensity decreases slower because of the more energy stored in the EM field (consequently a small fraction of the EM field energy is able to compensate for the local atomic dissipation). This is clear shown in Fig.~\ref{Sub2}(c), where the field intensity at $x_\-A+0^+$ is plotted. The larger slope when $t<T$ for the case of $\beta=0.5$ than the case of $\beta=0.2$ means a faster decay of the atoms. The slopes change at $t=2T$ because of the arrival of the atom B-reflected field that is emitted by atom A at $t=0$. This change is only obvious when $\beta$ is large \cite{Roy17RMP, Sheremet23RMP}. The dependence of effective decay rate $\Gamma$ on the retardation $T$ can be similarly considered, which determines how long the atoms decay independently before the establishment of the quasi-BS. 

\begin{figure}
    \centering
    \includegraphics[width=\columnwidth]{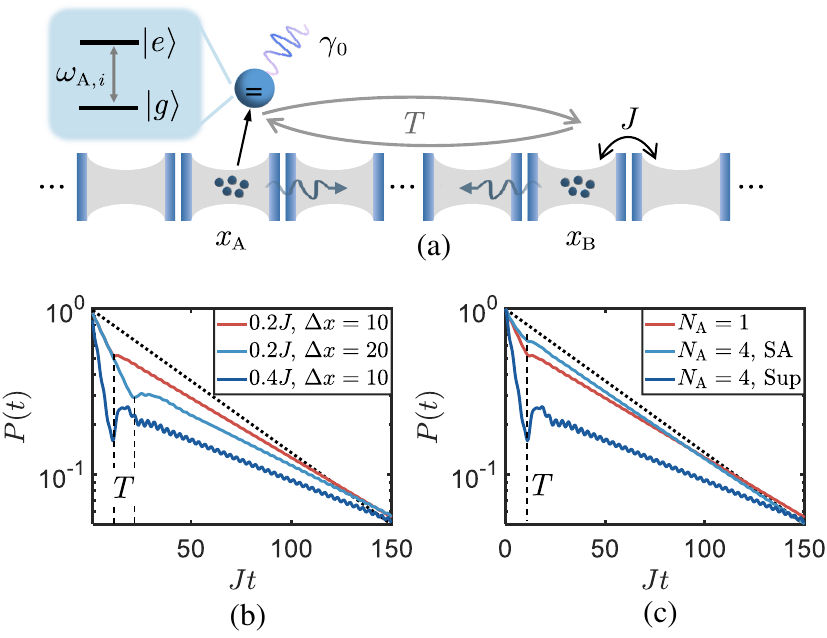}
    \caption{(a) Sketch of the waveguide consisting of tunnel-coupled cavities, where $J$ is the tunneling strength between neighboring cavities. Two collections of atoms are coupled to cavity $x_\mathrm{A}$ and cavity $x_\mathrm{B}$, each with $N_\mathrm{A}$ and $N_\mathrm{B}$ atoms, respectively.  $T$ sets a lower bound for a photon to have a round trip between $x_\mathrm{A}$ and $x_\mathrm{B}$. Atomic excitation probability $P(t)$ versus time (in units of $1/J$) are plotted in semi-logarithmic coordinate (b)~for different coupling strengths $g_\mathrm{A}=0.2J$, $0.4J$ or different inter-atom distances $\Delta x=10$, $20$ and (c)~for different number of atoms at $x_\mathrm{A}$. For $N_\mathrm{A}=4$, different initial state is considered, where there is only a single excited atom (denoted by ``SA'') or the initial state is in the superradiant state (denoted by ``Sup''). We set $N_\mathrm{A}=1$ in (b) and $g_\mathrm{A}=0.2J$, $\Delta x=10$ in (c), respectively, and $\sqrt{N_\mathrm{B}}g_\mathrm{B} =2J$ is kept unchanged in all cases. The black dotted line for exponential decay $P(t)=\exp(-\gamma_0t)$ is plotted for comparison.}
    \label{Sub3}
\end{figure}

\textit{Self-interference and cooperative emission.--} The BS can be established by a single excited atom which decays into a 1D semi-infinite waveguide with a mirror \cite{Tufarelli13PRA}. Specifically, the mirror can be replaced by an unexcited atom coupled to the waveguide \cite{Mirhosseini19N}. Here we study the quasi-BS in such an architecture in the non-Markovian regime. We show that the local decay rate can be suppressed by the non-Markovian retardation such that $\Gamma<\gamma_0$, and when we further replace the atom with a collection of atoms, the cooperative coupling results in a lower excited state decay rate of $\Gamma$.  

As displayed in Fig.~\ref{Sub3}(a), now we consider the waveguide formed by a 1D array of $N$ tunnel-coupled cavities \cite{Wang20May, Bello19SA, Scigliuzzo22PRX, Notomi08NPhoton}. The cavity array is described by a tight-binding Hamiltonian
\begin{equation}
    H_\-{ph}=\sum_{x=1}^N{\omega_\-c a_x\dagg a_x}-J\sum_{x=1}^{N-1}\left(a_{x+1}\dagg a_x+a_x\dagg a_{x+1}\right),
\end{equation}
where $a_x$ is the photon annihilation operator of cavity $x$ and $J>0$ is the tunneling strength. $\omega_c$ is the central frequency of each individual cavity and hereafter we mainly focus on the on-resonance situation where $\omega_c=\omega_\-A=\omega_\-B$, with $\omega_\-A$ and $\omega_\-B$ the transition frequencies of the two collection of atoms (each collection with $N_i$ atoms) in cavity $x_\-A$ and $x_\-B$, respectively. Then the total Hamiltonian reads
\begin{equation}
    H=H_\-{ph}+\sum_{\substack{i\in\{\mathrm{A,B}\}\\j\leq N_i}}\left[\:\omega_i\sigma_{i,j}^+\sigma_{i,j}+\left(g_ia_{x_i}\dagg \sigma_{i,j}+\-{H.c.}\right)\right].
\end{equation}
Here we include the local decay rate $\gamma_0$ by setting $\:\omega_i=\omega_i-\-i\gamma_0/2$. 

Instead of an entangled initial state, here we assume that only one atom in cavity A is excited at $t=0$. Without loss of generality we assume $\ket{\psi(0)}=\sigma_{\-A,1}^+\ket{G,\varnothing}$. At time $t$ the state has the following form
\begin{equation}
    \ket{\psi(t)}=\left(\sum_{i,j}c_{i,j}(t)\sigma_{i,j}^++\sum_xc_x(t)a_x\dagg\right)\ket{G,\varnothing},
\end{equation}
where $c_{i,j}(t)$ is the probability amplitude of atom $j$ in cavity A being excited and $c_x(t)$ the probability amplitude of one photon being located at cavity $x$. We calculate the evolution of the system directly by solving the Schr\"odinger's equation $\-i\partial_t\ket{\psi(t)}=H\ket{\psi(t)}$. The  excited state population $P(t)=\sum_j|c_{\-A,j}(t)|^2$ is presented in Fig.~\ref{Sub3}(b)(c). When $t<T$ (the exact value of $T$ will be specified later), $P(t)$ decays exponentially with rate $\gamma=\gamma_0+\gamma_{1\-D}$, with now $\gamma_{1\-D}=g_\-A^2/J$ \cite{Wang20May}. At time $t=T$, the reflected EM field begins to arrive, and the quasi-BS forms by \emph{self-interference}, thus reducing the total decay rate.

In order to evaluate the time scale needed for this delayed feedback \cite{Hein14Jul, Grimsmo15Aug, Pichler16Mar} to establish, the cavity modes are Fourier-transformed to the momentum representation ($k$-space) where $a_k=(1/\sqrt{N})\sum_xa_x\exp(\-ikx)$, $g_i^k=(1/\sqrt{N})g_i\exp(\-ikx_i)$ and $\omega_k=\omega_c-2J\cos k$ \cite{SM}. The waveguide dispersion relation gives rise to the peak group velocity $v_\-g^m=\max\{\-d\omega_k/\-dk\}=2J$ at $k_m=\pm\pi/2$. Consequently we have $T\sim 2|x_\-A-x_\-B|/v_\-g^m$, which must be spent before the reflected fields arrive. We set $d/2=|x_\-A-x_\-B|$ to be an even integer, i.e., $d/2=2n$, $n\in\mathbb{Z}$, which yields a propagating phase of $\sim |k_m|d=2n\pi$; combined with the $\pi$ shift induced by total reflection at cavity B,  the total accumulated phase in one round trip between A and B is $\phi=(2n+1)\pi$, thus leading to a destructive interference at cavity A \cite{Qiao22Aug}.

In Fig.~\ref{Sub3}(b), the dependence of $P(t)$ on different atom distances $\Delta x$ (consequently different retardation $T$) and different atom-cavity coupling strengths (consequently different coupling efficiency $\beta$) is displayed. The results can be similarly understood as that in Fig.~\ref{Sub1}(b)(c). Things are different when we consider a collection of atoms in the cavity. In Fig.~\ref{Sub3}(c), the results for $N_\-A=1, 4$ are presented. For $N_\-A=4$, if there is only a single excited atom (denoted by ``SA'') initially, the excited state decay rate $\Gamma(t)$ is suppressed when $t<T$ compared to the $N_\-A=1$ case, which stems from the reabsorption, i.e., the emitted photon can be absorbed by other atoms before tunneling into neighboring cavities. On the contrary, $\Gamma(t)$ is larger than the $N_\-A=1$ case when $t>T$, which we attribute to the reabsorption induced phase mismatch: destructive interference occurs between one atom at time $t$ and the photon emitted at $t-T$; when reabsorption exists, a photon takes longer time than $T$ for a round trip between A and B because of the time spent among atoms in cavity A. We thus prepare the initial state in the superradiant state (denoted by ``Sup'') $\ket{\psi(0)}=(1/2)\sum_{i=1}^{4}\sigma_{\-A,i}\ket{G,\varnothing}$. It is seen that the excited-state decay rate $\Gamma(t)$ is firstly improved when $t<T$ because of the local superradiance, and then suppressed when $t>T$ because of the improved coupling efficiency, as have been discussed in the two-atom case.

\begin{figure}
    \centering
    \includegraphics[width=0.67\columnwidth]{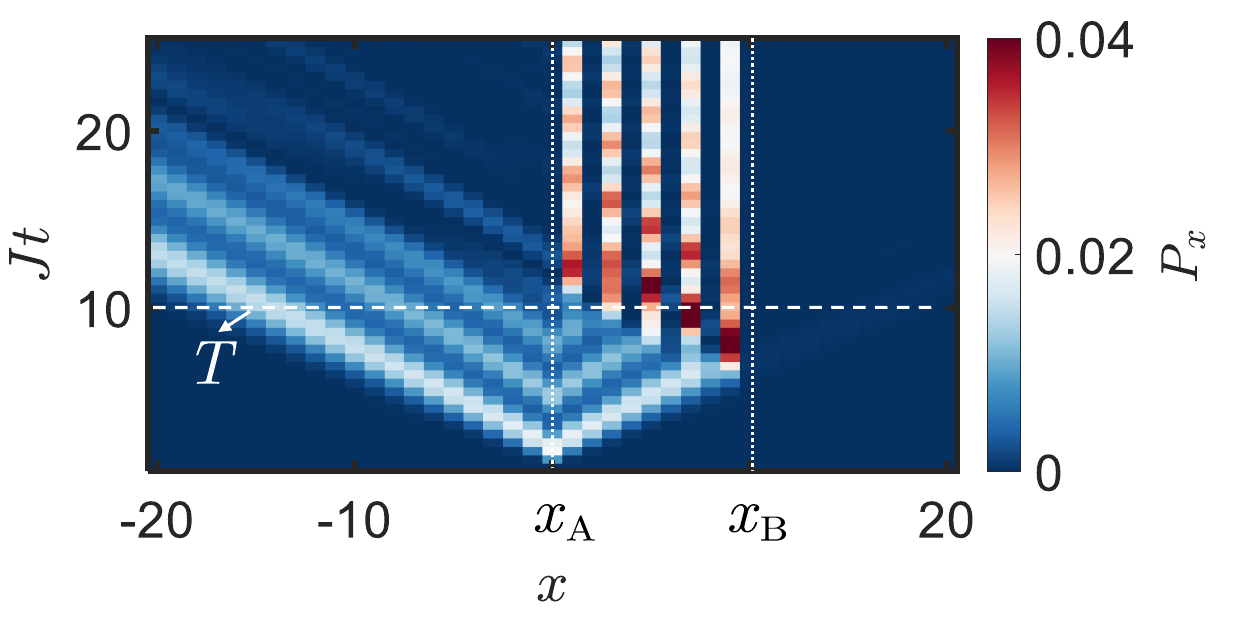}
    \caption{Photon distribution probability $P(x)=|c_x|^2$ as a function of position and time.}
    \label{Sub4}
\end{figure}

 Atoms in cavity B serve as a mirror \cite{Nie23Sep, Leonforte21Feb} and reflect the EM field. In all our calculations, we have kept $\sqrt{N_\-B}g_\-B=2J$ unchanged, as the reflection is determined by $\sqrt{N_\-B}g_\-B$ \cite{SM}, i.e., the number of atoms and the atom-cavity coupling strength.  When the reflected field arrives at cavity A, it interfere destructively with the real time emitted field and thus the EM fields are confined between $x_\-A$ and $x_\-B$, as displayed in Fig.~\ref{Sub4}. It is shown that after the right-propagating field arrives at $x_\-B$, the reflected field begins to interfere with the unreflected field to form a standing wave. Atoms at $x_\-A$ and $x_\-B$ are the nodes of this wave, and the system mimics the so called vacancy-like dressed states \cite{Nie23Sep, Leonforte21Feb}. The confined EM field compensates for the atomic dissipation similar to the case discussed in Fig.~\ref{Sub2}.

In summary, the non-Markovian subradiant state studied  in this work has a total decay rate smaller than the local decay rate of individual atoms, which is imparted by the retarded interaction and vanishes in the Markovian regime. Such decay-suppression effect can be boosted by increasing the inter-atom distance and the coupling efficiency, and occurs spontaneously in a self-interfering setup, and can be stressed by local superradiance. We anticipate that non-Markovian subradiant states could yield new error bounds and protocols for many applications in quantum technology, ranging from quantum memory to quantum metrology. The non-Markovian collective states could itself constitute a new realm with rich many-body physics.

\begin{acknowledgments}
  L.L. acknowledges support from the National Key Research and Development Program of China (grant 2020YFA0715000), the National Natural Science Foundation of China (grant 62075111), and  the Tsinghua University Initiative Scientific Research Program; H.-B.S. acknowledges support from the National Natural Science Foundation of China (grant 61960206003) and Tsinghua-Foshan Innovation Special Fund (grant 2021THFS0102).
\end{acknowledgments}
%
\end{document}